\documentclass{Interspeech2024}
\usepackage{multirow}
\usepackage{mathrsfs}

\interspeechcameraready

\title{CoLM-DSR: Leveraging Neural Codec Language Modeling for Multi-Modal Dysarthric Speech Reconstruction}

\name[affiliation={1}]{Xueyuan}{Chen}
\name[affiliation={1}]{Dongchao}{Yang}
\name[affiliation={1}]{Dingdong}{Wang}
\name[affiliation={1,3,*}]{Xixin}{Wu}
\name[affiliation={1,2,*}]{Zhiyong}{Wu}
\name[affiliation={1,2,3}]{Helen}{Meng}

\address{
  $^1$The Chinese University of Hong Kong, Hong Kong SAR, China\\
  $^2$Shenzhen International Graduate School, Tsinghua University, Shenzhen, China\\
  $^3$Vocal Engineering Technologies Limited, Hong Kong SAR, China}
\email{\{xychen, dcyang, ddwang, wuxx, zywu, hmmeng\}@se.cuhk.edu.hk\thanks{* Corresponding authors.}}


\begin{document}

\maketitle

\begin{abstract}
    Dysarthric speech reconstruction (DSR) aims to transform dysarthric speech into normal speech. It still suffers from low speaker similarity and poor prosody naturalness. In this paper, we propose a multi-modal DSR model by leveraging neural codec language modeling to improve the reconstruction results, especially for the speaker similarity and prosody naturalness. Our proposed model consists of: (i) a multi-modal content encoder to extract robust phoneme embeddings from dysarthric speech with auxiliary visual inputs; (ii) a speaker codec encoder to extract and normalize the speaker-aware codecs from the dysarthric speech, in order to provide original timbre and normal prosody; (iii) a codec language model based speech decoder to reconstruct the speech based on the extracted phoneme embeddings and normalized codecs. Evaluations on the commonly used UASpeech corpus show that our proposed model can achieve significant improvements in terms of speaker similarity and prosody naturalness\footnote[1]{\href{https://Chenxuey20.github.io/CoLM-DSR}{Audio samples: https://Chenxuey20.github.io/CoLM-DSR}}.
     
\end{abstract}

\section{Introduction}
Dysarthria is a prevalent type of speech disorder that is commonly observed in individuals with neuromotor conditions
such as Parkinson's disease and cerebral palsy \cite{darley1975motor,whitehill2000speech}.
This condition results in a significant deterioration in speech quality and voice characteristics from normal speech patterns \cite{darley1969differential},
which greatly hampers dysarthria patients' daily communication with their family members or caregivers \cite{kain2007improving}.
Dysarthric speech reconstruction (DSR) is a highly effective approach that seeks to improve the speech intelligibility, naturalness, and preserve the original speaker’s timbre by transforming dysarthric speech into normal speech.

The task of DSR is a complex endeavor that has garnered significant research attention.
The voice banking-based method collects pre-recorded normal speeches from dysarthric patients before their speech abilities deteriorate to develop personalized text-to-speech (TTS) systems \cite{yamagishi2012speech},
but its applicability is limited to individuals with available normal speech data \cite{chen2022hilvoice}.
The voice conversion (VC) based techniques aim to modify dysarthric speech signals to improve intelligibility and naturalness while preserving the content,
such as rule-based VC \cite{rudzicz2013adjusting,kumar2016improving}, and statistical VC approaches \cite{fu2016joint,aihara2017phoneme}.
Recently, an end-to-end VC based DSR system \cite{wang2020end} has been proposed, 
which involves distilling a speech encoder from a pre-trained automatic speech recognition (ASR) model to replace the text encoder in a sequence-to-sequence (seq2seq) TTS system. 
Compared to a cascaded system that relies on ASR results for TTS, it does not restrict intermediate representations to text characters and can generate speech with lower errors and higher fidelity. 
Motivated by the prosody and timbre modeling in TTS systems \cite{chen2022character,chen2022unsupervised,chen2024stylespeech}, additional components, such as a prosody corrector and speaker encoder, have been introduced to further enhance prosody and speaker similarity \cite{wang2020learning}.
To improve the speech intelligence for patients with severe dysarthria, as well as for speech captured from complex, noisy acoustic environments, a multi-modal framework \cite{chen2024exploiting} is first proposed.
Two multi-modal encoders are designed and compared to utilize visual information
, e.g., lip movements, 
as additional clues for reconstructing the highly abnormal pronunciations.
In order to address the issue of training inefficiency due to complex training strategies and cascaded pipelines, Unit-DSR \cite{wang2024unit} is proposed to use the discrete speech units extracted from HuBERT \cite{hsu2021hubert} for the generation of a normal speech waveform. 

Though significant progress has been made, most existing works has focused primarily on improving the speech intelligibility \cite{wang2020end,chen2024exploiting,wang2024unit}.
However, the speaker similarity and prosody naturalness, which are also crucial to a patient's sense of self-identity and fluent expression, still leave a lot to be desired.
In most real-world application scenarios, it is crucial for DSR models to exhibit quick adaptation abilities to new dysarthric patients with limited data,
which is difficult for existing speaker encoder based DSR systems \cite{wang2020learning}.
With the development of advanced prompting-based language model (LM) in the field of text analysis \cite{brown2020language} and 
audio processing \cite{borsos2023audiolm},
some zero-shot TTS frameworks have shown strong in-context learning capabilities and diverse outputs with improved speaker similarity and speech naturalness \cite{wang2023neural, zhang2023speak, ji2023textrolspeech},
which treat TTS as a language model task with audio codecs as an intermediate representation instead of the traditional mel-spectrogram.

Inspired by the success of neural codec language modeling in zero-shot TTS \cite{wang2023neural}, 
this paper proposes a codec LM based multi-modal DSR system by leveraging neural codec language modeling 
with the large, diverse, and multi-speaker normal speech data
to improve the reconstruction results, especially for the speaker similarity and prosody naturalness.
Firstly, a multi-modal content encoder is adopted to extract the robust phoneme embeddings from dysarthric speech with auxiliary visual inputs.
Secondly, we design a speaker codec encoder to extract and modify the speaker-aware codecs from the dysarthric speech, in order to provide the acoustic prompts with original timbre and normal prosody.
Thirdly, we use the speech decoder by leverage the neural codec language model to generate the reconstructed speech based on the extracted phoneme embeddings and normal speaker-aware codecs.
The contributions of this paper include:
\begin{itemize}
\item We propose the first codec LM based multi-modal DSR system by combining an audio-visual encoder with the neural codec LM framework to reconstruct the dysarthric speech.

\item We specially design a novel speaker codec encoder for DSR task by mapping the dysarthric codecs into normal codecs to provide original timbre and normal prosody prompts.

\item Both subjective and objective experimental results show that our proposed codec LM based DSR system achieves significant improvements especially for the speaker similarity and prosody naturalness.
\end{itemize}

\begin{figure}[t]
\centering
\includegraphics[width=0.94\columnwidth]{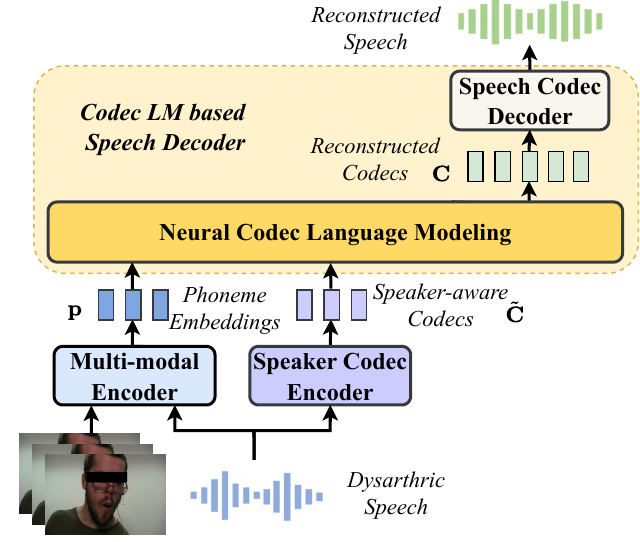}
\vspace{-5pt}
\caption{Overall architecture of the proposed CoLM-DSR system with Multi-modal Content Encoder, Speaker Codec Encoder and Codec LM based Speech Decoder.}
\label{fig:model_overall}
\vspace{-15pt}
\end{figure}

\begin{figure*}[ht]
\centering
\includegraphics[width=2.10\columnwidth]{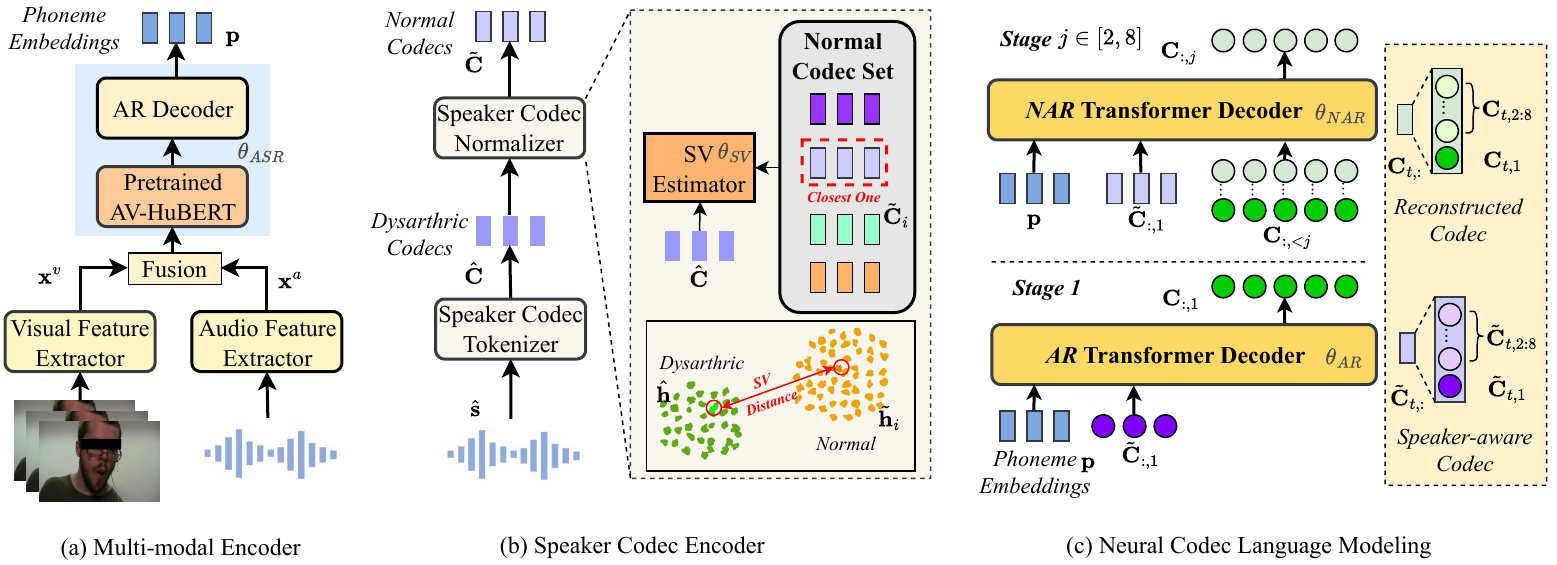}
\vspace{-7pt}
\caption{Diagram of the proposed CoLM-DSR system, where (a), (b) and (c) show the model details of Multi-modal Encoder, Speech Codec Encoder and Neural Codec Language Modeling respectively.}
\vspace{-10pt}
\label{fig:model_details}
\end{figure*}

\vspace{-5pt}
\section{Methodology}
Our proposed CoLM-DSR model is illustrated in Figure \ref{fig:model_overall}.
It mainly consists of a multi-modal content encoder, a speaker codec encoder and a codec LM based speech decoder.
Specifically, the multi-modal content encoder strives to extract robust phoneme embeddings from dysarthric audio and visual inputs to provide content prompts.
The speaker codec encoder is designed to extract and normalize the speaker-aware codecs from dysarthric speech to provide timbre and prosody prompts.
The speech decoder takes phoneme embeddings and speaker-aware codecs as prompt inputs to generate the reconstructed speech.

\subsection{Multi-modal Encoder for Content Extraction}
\label{sec:audio-visual speech encoder}
To reconstruct the linguistic content of original dysarthric speech, a multi-modal encoder is used to extract robust linguistic representations.
Following \cite{chen2024exploiting}, we adopt the multi-modal encoder outputs, i.e., the phoneme probability distribution, as phoneme embeddings, which are denoted as $\mathbf{p}$.

As shown in Figure \ref{fig:model_details} (a), the multi-modal encoder contains an audio feature extractor, a visual feature extractor and an auto-regressive (AR) seq2seq ASR model.
Firstly, the log-MMSE speech enhancement
algorithm \cite{ephraim1985speech} is adopted to reduce the strong background noise, and then 80-dimension filter banks (FBKs)+$\Delta$ features are extracted as the audio features $\mathbf{x}^a$.
Secondly,
an off-the-shelf
face alignment network \cite{bulat2017far} is employed to detect the lip landmarks,
followed by discrete cosine transform (DCT) and linear discriminant analysis (LDA) to downsize
and obtain the visual features $\mathbf{x}^v$.
After that, we further take the common operation of concatenating the temporally aligned audio and visual features along the feature dimension,
which is followed by a fully-connected (FC) layer to fuse and get the audio-visual features.
The audio-visual features are further fed into the following seq2seq ASR model $\theta_{ASR}$.
It consists of a pretrained AV-HuBERT transformer encoder \cite{shi2022avhubert} and an AR decoder with connectionist temporal classification (CTC) \cite{liu2019adversarial}.
The decoder contains a 512-dimensional location-aware attention module and 2 LSTM layers with 1024 units per layer.
Finally, the ASR model outputs the phoneme embeddings $\mathbf{p}$,
which can be described as 
\begin{equation}
    \begin{split}
        & \mathbf{p}=f_{ASR}(\mathbf{W}_1 (\mathbf{x}^a\oplus\mathbf{x}^v)+\mathbf{z}_1;\theta_{ASR})
    \end{split}
\end{equation}
where $\oplus$
is concatenation along the feature dimension, $\mathbf{W}_1$ and $\mathbf{z}_1$ are FC-layer parameters.

\subsection{Speaker Codec Encoder for Timbre Preservation and Prosody Normalization}
We specially design a speaker codec encoder by mapping the dysarthric codecs into normal codecs to provide speaker-aware timbre and prosody prompts.
As shown in Figure \ref{fig:model_details} (b), our proposed speaker codec encoder consists of two modules: speaker codec tokenizer and speaker codec normalizer.

\textbf{Speaker Codec Tokenizer:}
To be specific, we adopt a pre-trained neural audio codec model, EnCodec \cite{defossez2022high}, as our tokenizer.
EnCodec is a convolutional encoder-decoder model, whose input and output are both 24 kHz audio across variable bitrates.
Each embedding produced by the EnCodec encoder is modeled by a residual vector quantization (RVQ) \cite{vasuki2006review}, in which the \textbf{8} hierarchy quantizers with 1024 entries are finally chosen.
Therefore, for each given dysarthric speech $\mathbf{\hat{s}}$ or normal speech $\mathbf{\tilde{s}}$, the corresponding 
codecs can be obtained and denoted as: 
\begin{equation}
    \begin{split}
        & \hat{\mathbf{C}}^{T\times8}=EnCodec(\hat{s}), \tilde{\mathbf{C}}^{T\times8}=EnCodec(\tilde{s}),
    \end{split}
\end{equation}
where $\hat{\mathbf{C}}$ and $\tilde{\mathbf{C}}$  represent the two-dimensional acoustic code matrix, and $T$ is the downsampled utterance length.

\textbf{Speaker Codec Normalizer:}
Since the dysarthric codecs contain abnormal prosodic features and severe noise information, their distribution is different from that of normal codecs,
which will seriously affect the prosody and sound quality of reconstructed speech.
Therefore, we further design a speaker codec normalizer to map the dysarthric codecs $\hat{\mathbf{C}}$ into corresponding normal codecs $\Tilde{\mathbf{C}}$ by a speaker verification (SV) estimator $\theta_{SV}$ to not only preserve the timbre but also modify the prosody.
The SV estimator is trained with a generalized end-to-end (GE2E) loss \cite{wan2018generalized},
so that the hidden codec representations $\mathbf{h}$ extracted from codec sequence $\mathbf{C}$ of the same speaker and different speakers have high and low similarity, respectively.
By utilizing a substantial collection of high-quality speech recordings from diverse speakers with varying timbres and natural prosody, we develop a comprehensive normal codec set $\mathscr{C}=\{\tilde{\mathbf{C}}_i : i=1, ... , N\}$,
which can be considered as encompassing the entire space within the domain of normal codecs.
Based on this, for any given dysarthric codec sequence $\hat{\mathbf{C}}$, we can map it to the closest one $\tilde{\mathbf{C}}$ in the normal codec set $\mathscr{C}$ by SV distance,
where the SV distance is the L1 distance between the hidden codec representations $\hat{\mathbf{h}}$ and $\tilde{\mathbf{h}}_i$ extracted by the SV estimator $\theta_{SV}$, formulated as:
\begin{equation}
    \begin{split}
        & \hat{\mathbf{h}}_{\sim\hat{\mathbf{C}}}=f_{SV}(\hat{\mathbf{C}};\theta_{SV}), \tilde{\mathbf{h}}_{i\sim\tilde{\mathbf{C}}_i}=f_{SV}(\tilde{\mathbf{C}_i};\theta_{SV})
    \end{split}
\end{equation}
\begin{equation}
    \begin{split}
        & \hat{\mathbf{C}} \to \tilde{\mathbf{C}} = \arg\min_{\hat{\mathbf{C}}_i \in \mathscr{C}} {\mid \hat{\mathbf{h}}_{\sim\hat{\mathbf{C}}} - \tilde{\mathbf{h}}_{i\sim\tilde{\mathbf{C}}_i} \mid}
    \end{split}
\end{equation}
After
that, the corresponding normal speaker-aware codecs $\tilde{\mathbf{C}}$ with original timbre and natural prosody are obtained.

\subsection{Codec LM based Speech Decoder for Speech Reconstruction}
Inspired by the zero-shot TTS \cite{wang2023neural}, we leverage the neural codec LM based decoder to reconstruct the dysarthric speech conditioned on the phoneme embeddings $\mathbf{p}$ and speaker-aware codecs $\tilde{\mathbf{C}}$.
The neural codec LM is expected to learn to extract the content and speaker information from the phoneme embeddings and the codecs respectively.
Figure \ref{fig:model_details} (c) shows the process of neural codec language modeling.
Specifically, there are two conditional language models in a hierarchical manner.

\textbf{AR Transformer Decoder:} During stage 1, an autoregressive (AR) transformer decoder $\theta_{AR}$ is adopted for the first quantizer $\mathbf{C}_{:, 1}$,
which is conditioned on the phoneme embeddings $\mathbf{p}$ and the first quantizer of acoustic codecs $\tilde{\mathbf{C}}_{:, 1}$, formulated as
\begin{equation}
    \begin{split}
        &p(\mathbf{C}_{:, 1} | \mathbf{p}, \tilde{\mathbf{C}}_{:, 1} ; \theta_{A R})=\prod_{t=0}^{T} p(\mathbf{C}_{t, 1} | \mathbf{C}_{<t, 1}, \tilde{\mathbf{C}}_{:, 1}, \mathbf{p} ; \theta_{A R})
    \end{split}
\end{equation}

\textbf{NAR Transformer Decoder:} After obtaining the first quantizer codecs $\tilde{\mathbf{C}}_{:, 1}$ by the AR model and during stage 2-8, a non-autogressive (NAR) transformer decoder $\theta_{NAR}$ is used for generating the discrete codecs from the second to the last quantizers, $\mathbf{C}_{:,j\in[2,8]}$.
It is conditioned on the phoneme embeddings  $\mathbf{p}$, the speaker-aware codecs $\tilde{\mathbf{C}}$ and the predicted acoustic tokens belong to the previous codebooks $\mathbf{C}_{:, <j}$: 
\begin{equation}
    \begin{split}
        &p(\mathbf{C}_{:,2:8}| \mathbf{p}, \tilde{\mathbf{C}} ; \theta_{NAR})=\prod_{j=2}^{8} p(\mathbf{C}_{:, j}| \mathbf{C}_{:, <j},\mathbf{p}, \tilde{\mathbf{C}}; \theta_{NAR})
    \end{split}
\end{equation}

Finally, the whole reconstructed codecs $\mathbf{C}=\mathbf{C}_{:,1}\oplus\mathbf{C}_{:,2:8}$ with 8 quantizer codes are obtained by the concatenation of each stage result.
Then the pre-trained speech codec decoder \cite{defossez2022high} is used to synthesize the reconstructed speech.

\section{Experiments}
\subsection{Experimental Settings}
Experiments are conducted on the UASpeech \cite{kim2008dysarthric}, VCTK \cite{veaux2016superseded} and LibriTTS \cite{zen2019libritts} datasets.
The UASpeech corpus is a benchmark disordered speech corpus, which is recorded by an 8-channel microphone array and a video camera with some background noise.
We use the VCTK corpus with 105 native speakers to train the SV estimator.
The LibriTTS corpus containing 580 hours of  normal speech from 2456 speakers is used to 
develop the normal codec set and 
train the Codec LM based speech decoder by teacher-forcing mode.
Similar to \cite{chen2024exploiting}, four speaker-dependent DSR systems are separately built for the four selected speakers (M12, F02, M16 and F04) with the lowest speech intelligibility.
Three mel-spectrogram based baseline settings are compared:

\begin{itemize}
\item \textbf{AON-DSR}: It uses an audio-only encoder to extract phoneme embeddings, a prosody corrector to explicitly model the duration and pitch, a speaker encoder to represent the speaker embedding, and a mel-decoder to reconstruct the mel-spectrogram based on the phoneme and prosody inputs\cite{wang2020learning}.

\item \textbf{VGG-DSR}: Following AON-DSR system, it uses a VGG-based audio-visual encoder instead of the audio-only encoder to extract phoneme embeddings \cite{chen2024exploiting}.

\item \textbf{AVHu-DSR}: Following AON-DSR system, it uses a AVHu-\\BERT-based audio-visual encoder (similar to \ref{sec:audio-visual speech encoder}) instead of the audio-only encoder to extract phoneme embeddings \cite{chen2024exploiting}.


\end{itemize}

All the content encoders are first trained on the whole dysarthric speech datasets of all speakers for 1M steps with batch size of 8, and then finetuned on the target speaker for 2k steps to improve phoneme prediction accuracy.
The open-source pre-trained `AV-HuBERT Base' model\footnote[1]{\href{https://github.com/facebookresearch/av_hubert}{https://github.com/facebookresearch/av\_hubert}} 
and `EnCodec' model\footnote[2]{\href{https://github.com/facebookresearch/encodec}{https://github.com/facebookresearch/encodec}}
are adopted in our experiments.
The codec LM based decoder is implemented based on an open-source implementation \footnote[3] {\href{https://github.com/lifeiteng/vall-e}{https://github.com/lifeiteng/vall-e}} of VALL-E \cite{wang2023neural},
and is trained on 4 NVIDIA V100 GPUs for 300K iterations with a batch size of 4 on each GPU.

\begin{table*}[!htp]
\caption{Comparison Results of MOS with 95\% Confidence Intervals for Speaker Similarity and Speech Naturalness.}
\vspace{-7pt}
\label{tab:subjective}
\centering
\begin{tabular}{c|cccc|cccc}
\toprule
\multirow{2}{*}{Models}  & \multicolumn{4}{c|}{Speaker Similarity} &\multicolumn{4}{c}{Speech Naturalness}    \\ 
& M12       & F02  & M16   & F04 & M12 & F02 & M16   & F04   \\ 
\hline
Original  & -  & -    & -   &  -   & 1.44$\pm$0.19  & 1.70$\pm$0.17    & 2.44$\pm$0.22   &  2.68$\pm$0.25 \\
AON-DSR  & 1.90$\pm$0.21  & 2.61$\pm$0.11    & 2.54$\pm$0.18   &  2.81$\pm$0.11   & 2.90$\pm$0.33  & 2.78$\pm$0.31    & 2.98$\pm$0.32   &  3.12$\pm$0.36 \\
VGG-DSR  & 2.32$\pm$0.18  & 2.78$\pm$0.17    & 2.66$\pm$0.21   &  2.92$\pm$0.13   & 3.00$\pm$0.32  & 2.94$\pm$0.29    & 3.26$\pm$0.28   & 3.20$\pm$0.33 \\
AVHu-DSR  & 2.31$\pm$0.21  & 2.81$\pm$0.15    & 2.69$\pm$0.19   &  3.10$\pm$0.20   & 3.56$\pm$0.24  & 3.62$\pm$0.19    & 3.54$\pm$0.19   &  3.52$\pm$0.26 \\
Proposed  & \textbf{3.30$\pm$0.17}  & \textbf{3.70$\pm$0.22}    & \textbf{3.58$\pm$0.18}   &  \textbf{3.78$\pm$0.19}   & \textbf{3.90$\pm$0.33}  & \textbf{3.80$\pm$0.32}    & \textbf{3.80$\pm$0.33}   &  \textbf{3.91$\pm$0.22} \\
\bottomrule
\end{tabular}
\vspace{-10pt}
\end{table*}

\subsection{Experimental Results}
\subsubsection{Speaker Similarity Comparison}
Subjective tests are conducted to evaluate the speaker similarity of reconstructed speech compared with the original dysarthric speech.
10 subjects are invited to give the 5-point mean opinion score (MOS, 1-bad, 2-poor, 3-fair, 4-good, 5-excellent) for 10 utterances randomly selected from each of four dysarthric speakers, and the scores are averaged and shown in Table \ref{tab:subjective}.
As can be observed, all baseline systems still have a poor speaker similarity performance.
Compared with the three speaker encoder based baseline systems, our proposed codec LM based model achieves significant improvements for all the 4 speakers on speaker similarity.

We also employ the speaker verification model \cite{wan2018generalized} as an objective measure to evaluate the speaker similarity between the dysarthric speeches and corresponding reconstructed speeches, and results are shown in Table \ref{tab:objective speaker similarity}.
Our proposed model also achieves the best results for all speakers.
Both the subjective and objective results illustrate that our proposed codec prompting based DSR system can preserve more original timbre information benefited from the zero-shot voice clone ability of codec LM.
Compared with the speaker encoder based baseline methods requiring large data, our codec LM based model is more suitable for this low-source DSR task.

\vspace{-3pt}
\begin{table}[ht]
\caption{Objective Comparison Results for Speaker Similarity.}
\vspace{-7pt}
\label{tab:objective speaker similarity}
\centering
\begin{tabular}{ccccc}
\toprule
Models   & M12  & F02  & M16  & F04 \\ \hline
AON-DSR    & 1.135  & 1.161 & 1.070 & 1.061      \\ 
VGG-DSR  & 1.137  & 1.152 & 1.066 & 1.069     \\ 
AVHu-DSR  & 1.127  & 1.154 & 1.074 & 1.054     \\ 
Proposed  & \textbf{1.080}  & \textbf{1.074} & \textbf{0.987} & \textbf{0.969}     \\ 
\bottomrule
\end{tabular}
\end{table}
\vspace{-10pt}

\subsubsection{Speech Naturalness Comparison}
To show the speech naturalness improvement of final reconstructed speech compared with the 'Original' dysarthric speech, we also conduct a MOS test on prosody performance.
As shown in Table \ref{tab:subjective},
we can see that the original dysarthric speech obtains the lowest score and suffers from very severely abnormal prosody.
All DSR systems improve the naturalness of original dysarthric speech, 
and our proposed CoLM-DSR system achieves the highest scores for all dysarthric patients.
The baseline systems rely on predicting the explicit prosodic features (e.g., duration and pitch) from phoneme embeddings during inference,
which tends to reconstruct an averaged prosody representation.
In contrast, our CoLM-DSR system is trained with large and diverse data in the LM manner,
and reconstructed directly based on the acoustic codec prompt during inference,
which can generate more natural and diverse prosody effects.

\subsubsection{Speech Intelligibility Comparison}
To show the content intelligibility of final reconstructed speech compared with 'Original' dysarthric speech, we use the publicly released ASR model, Whisper \cite{radford2023robust}, to obtain the word error rate (WER) with greedy decoding. The results are shown in Table \ref{tab:objective speech intelligibility}.
Compared with the original dysarthric speech, all the DSR systems achieve a significant improvement.
Our proposed CoLM-DSR system also achieves the best results for all speakers.
Compared with the SOTA baseline system AVHu-DSR,
the WER improvement is not obvious enough, since we adopt the same AVHuBERT-based encoder to extract the content information.
It also shows the content encoder is quite important for the speech intelligibility of reconstructed speech.

\vspace{-3pt}
\begin{table}[ht]
\caption{WER Comparison Results for Speech Intelligibility.}
\vspace{-7pt}
\label{tab:objective speech intelligibility}
\centering
\begin{tabular}{ccccc}
\toprule
Models   & M12  & F02  & M16  & F04 \\ \hline
Original   & 98.0\%  & 93.4\% & 81.6\% & 67.6\%   \\ 
AON-DSR    & 65.7\%  & 62.5\% & 60.3\% & 55.6\%      \\ 
VGG-DSR  & 60.6\%  & 59.2\% & 57.0\% & 55.4\%     \\ 
AVHu-DSR  & 55.9\%  & 56.1\% & 51.3\% & 48.0\%     \\ 
Proposed  & \textbf{55.6\%}  & \textbf{55.9\%} & \textbf{51.0\%} & \textbf{47.9\%}     \\ 
\bottomrule
\end{tabular}
\end{table}
\vspace{-10pt}

\subsubsection{Investigation on Dysarthric Codecs and Normal Codecs}
\begin{figure}[htp]
\centering
\includegraphics[width=0.85\columnwidth]{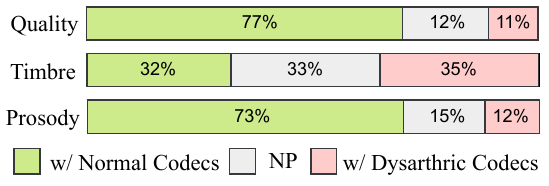}
\vspace{-7pt}
\caption{Results of AB Preference Test.}
\vspace{-10pt}
\label{fig:abx}
\end{figure}

In order to verify the necessity and effectiveness of our proposed speaker codec normalizer, we further conduct an analysis of the dysarthric codecs and normal codecs.
We use the dysarthric codecs as the acoustic prompts directly to generate the reconstructed speech.
However, we find that it is difficult to generate normal speech using dysarthric codecs directly for the severe patients, such as M12.
Therefore, we only select F04 as an example for comparison.
We perform three AB preference tests in terms of audio quality, timbre similarity and prosody naturalness respectively.
Listeners are required to select a better utterance for each given utterance pair.
The results are shown in Figure \ref{fig:abx}.
We can observe that using normal codecs and using dysarthric codecs achieve relatively consistent performance on timbre similarity.
While in terms of audio quality and prosody naturalness, the effect of using normal codecs is significantly better than that of using dysarthric codecs.
The results show that abnormal prosodic features and severe noise information seriously affect the prosody naturalness and audio quality of reconstructed speech,
and our proposed speaker codec normalizer can effectively not only preserve the original timbre but also improve the prosody and audio quality performance.

\vspace{-5pt}

\section{Conclusion}

This paper proposes to leverage the neural codec language model for improving dysarthric speech reconstruction results.
We combine an audio-visual content encoder with the neural codec language modeling framework.
To provide original speaker timbre and natural prosody acoustic prompts, we specially design a normal speaker codec encoder with codec tokenizer and normalizer by mapping the dysarthric codes to normal codecs.
Both subjective and objective experimental results on UASpeech corpus show that our proposed CoLM-DSR system can achieve significant improvements especially in terms of speaker similarity and prosody naturalness. 

\section{Acknowledgements}
This research is supported by National Natural Science Foundation of China (62076144), Shenzhen Science and Technology Program (WDZC20220816140515001, JCYJ20220818101014\\030), CUHK Direct Grant for Research (Ref. No. 4055221), the CUHK Stanley Ho Big Data Decision Analytics Research Centre and the Centre for Perceptual and Interactive Intelligence.


\bibliographystyle{IEEEtran}
\bibliography{mybib}

\begin{thebibliography}{10}
\providecommand{\url}[1]{#1}
\csname url@samestyle\endcsname
\providecommand{\newblock}{\relax}
\providecommand{\bibinfo}[2]{#2}
\providecommand{\BIBentrySTDinterwordspacing}{\spaceskip=0pt\relax}
\providecommand{\BIBentryALTinterwordstretchfactor}{4}
\providecommand{\BIBentryALTinterwordspacing}{\spaceskip=\fontdimen2\font plus
\BIBentryALTinterwordstretchfactor\fontdimen3\font minus \fontdimen4\font\relax}
\providecommand{\BIBforeignlanguage}[2]{{%
\expandafter\ifx\csname l@#1\endcsname\relax
\typeout{** WARNING: IEEEtran.bst: No hyphenation pattern has been}%
\typeout{** loaded for the language `#1'. Using the pattern for}%
\typeout{** the default language instead.}%
\else
\language=\csname l@#1\endcsname
\fi
#2}}
\providecommand{\BIBdecl}{\relax}
\BIBdecl

\bibitem{darley1975motor}
F.~L. Darley, A.~E. Aronson, and J.~R. Brown, ``Motor speech disorders,'' \emph{(No Title)}, 1975.

\bibitem{whitehill2000speech}
T.~L. Whitehill and V.~Ciocca, ``Speech errors in cantonese speaking adults with cerebral palsy,'' \emph{Clinical linguistics \& phonetics}, vol.~14, no.~2, pp. 111--130, 2000.

\bibitem{darley1969differential}
F.~L. Darley, A.~E. Aronson, and J.~R. Brown, ``Differential diagnostic patterns of dysarthria,'' \emph{Journal of speech and hearing research}, vol.~12, no.~2, pp. 246--269, 1969.

\bibitem{kain2007improving}
A.~B. Kain, J.-P. Hosom, X.~Niu, J.~P. Van~Santen, M.~Fried-Oken, and J.~Staehely, ``Improving the intelligibility of dysarthric speech,'' \emph{Speech communication}, vol.~49, no.~9, pp. 743--759, 2007.

\bibitem{yamagishi2012speech}
J.~Yamagishi, C.~Veaux, S.~King, and S.~Renals, ``Speech synthesis technologies for individuals with vocal disabilities: Voice banking and reconstruction,'' \emph{Acoustical Science and Technology}, vol.~33, no.~1, pp. 1--5, 2012.

\bibitem{chen2022hilvoice}
X.~Chen, Q.~Huang, X.~Wu, Z.~Wu, and H.~Meng, ``Hilvoice: Human-in-the-loop style selection for elder-facing speech synthesis,'' in \emph{ISCSLP}.\hskip 1em plus 0.5em minus 0.4em\relax IEEE, 2022, pp. 86--90.

\bibitem{rudzicz2013adjusting}
F.~Rudzicz, ``Adjusting dysarthric speech signals to be more intelligible,'' \emph{Computer Speech \& Language}, vol.~27, no.~6, pp. 1163--1177, 2013.

\bibitem{kumar2016improving}
S.~A. Kumar and C.~S. Kumar, ``Improving the intelligibility of dysarthric speech towards enhancing the effectiveness of speech therapy,'' in \emph{2016 International Conference on Advances in Computing, Communications and Informatics (ICACCI)}.\hskip 1em plus 0.5em minus 0.4em\relax IEEE, 2016, pp. 1000--1005.

\bibitem{fu2016joint}
S.-W. Fu, P.-C. Li, Y.-H. Lai, C.-C. Yang, L.-C. Hsieh, and Y.~Tsao, ``Joint dictionary learning-based non-negative matrix factorization for voice conversion to improve speech intelligibility after oral surgery,'' \emph{IEEE Transactions on Biomedical Engineering}, vol.~64, no.~11, pp. 2584--2594, 2016.

\bibitem{aihara2017phoneme}
R.~Aihara, T.~Takiguchi, and Y.~Ariki, ``Phoneme-discriminative features for dysarthric speech conversion.'' in \emph{Interspeech}, 2017, pp. 3374--3378.

\bibitem{wang2020end}
D.~Wang, J.~Yu, X.~Wu, S.~Liu, L.~Sun, X.~Liu, and H.~Meng, ``End-to-end voice conversion via cross-modal knowledge distillation for dysarthric speech reconstruction,'' in \emph{ICASSP 2020}.\hskip 1em plus 0.5em minus 0.4em\relax IEEE, 2020, pp. 7744--7748.

\bibitem{chen2022character}
X.~Chen, C.~Song, Y.~Zhou, Z.~Wu, C.~Chen, Z.~Wu, and H.~Meng, ``A character-level span-based model for mandarin prosodic structure prediction,'' in \emph{ICASSP 2022}.\hskip 1em plus 0.5em minus 0.4em\relax IEEE, 2022, pp. 7602--7606.

\bibitem{chen2022unsupervised}
X.~Chen, S.~Lei, Z.~Wu, D.~Xu, W.~Zhao, and H.~Meng, ``Unsupervised multi-scale expressive speaking style modeling with hierarchical context information for audiobook speech synthesis,'' in \emph{COLING}, 2022, pp. 7193--7202.

\bibitem{chen2024stylespeech}
X.~Chen, X.~Wang, S.~Zhang, L.~He, Z.~Wu, X.~Wu, and H.~Meng, ``Stylespeech: Self-supervised style enhancing with vq-vae-based pre-training for expressive audiobook speech synthesis,'' in \emph{ICASSP 2024}.\hskip 1em plus 0.5em minus 0.4em\relax IEEE, 2024, pp. 12\,316--12\,320.

\bibitem{wang2020learning}
D.~Wang, S.~Liu, L.~Sun, X.~Wu, X.~Liu, and H.~Meng, ``Learning explicit prosody models and deep speaker embeddings for atypical voice conversion,'' \emph{arXiv preprint arXiv:2011.01678}, 2020.

\bibitem{chen2024exploiting}
X.~Chen, Y.~Wang, X.~Wu, D.~Wang, Z.~Wu, X.~Liu, and H.~Meng, ``Exploiting audio-visual features with pretrained av-hubert for multi-modal dysarthric speech reconstruction,'' \emph{arXiv preprint arXiv:2401.17796}, 2024.

\bibitem{wang2024unit}
Y.~Wang, X.~Wu, D.~Wang, L.~Meng, and H.~Meng, ``Unit-dsr: Dysarthric speech reconstruction system using speech unit normalization,'' \emph{arXiv preprint arXiv:2401.14664}, 2024.

\bibitem{hsu2021hubert}
W.-N. Hsu, B.~Bolte, Y.-H.~H. Tsai, K.~Lakhotia, R.~Salakhutdinov, and A.~Mohamed, ``Hubert: Self-supervised speech representation learning by masked prediction of hidden units,'' \emph{IEEE/ACM Transactions on Audio, Speech, and Language Processing}, vol.~29, pp. 3451--3460, 2021.

\bibitem{brown2020language}
T.~Brown, B.~Mann, N.~Ryder, M.~Subbiah, J.~D. Kaplan, P.~Dhariwal, A.~Neelakantan, P.~Shyam, G.~Sastry, A.~Askell \emph{et~al.}, ``Language models are few-shot learners,'' \emph{Advances in neural information processing systems}, vol.~33, pp. 1877--1901, 2020.

\bibitem{borsos2023audiolm}
Z.~Borsos, R.~Marinier, D.~Vincent, E.~Kharitonov, O.~Pietquin, M.~Sharifi, D.~Roblek, O.~Teboul, D.~Grangier, M.~Tagliasacchi \emph{et~al.}, ``Audiolm: a language modeling approach to audio generation,'' \emph{IEEE/ACM Transactions on Audio, Speech, and Language Processing}, 2023.

\bibitem{wang2023neural}
C.~Wang, S.~Chen, Y.~Wu, Z.~Zhang, L.~Zhou, S.~Liu, Z.~Chen, Y.~Liu, H.~Wang, J.~Li \emph{et~al.}, ``Neural codec language models are zero-shot text to speech synthesizers,'' \emph{arXiv preprint arXiv:2301.02111}, 2023.

\bibitem{zhang2023speak}
Z.~Zhang, L.~Zhou, C.~Wang, S.~Chen, Y.~Wu, S.~Liu, Z.~Chen, Y.~Liu, H.~Wang, J.~Li \emph{et~al.}, ``Speak foreign languages with your own voice: Cross-lingual neural codec language modeling,'' \emph{arXiv preprint arXiv:2303.03926}, 2023.

\bibitem{ji2023textrolspeech}
S.~Ji, J.~Zuo, M.~Fang, Z.~Jiang, F.~Chen, X.~Duan, B.~Huai, and Z.~Zhao, ``Textrolspeech: A text style control speech corpus with codec language text-to-speech models,'' \emph{arXiv preprint arXiv:2308.14430}, 2023.

\bibitem{ephraim1985speech}
Y.~Ephraim and D.~Malah, ``Speech enhancement using a minimum mean-square error log-spectral amplitude estimator,'' \emph{IEEE transactions on acoustics, speech, and signal processing}, vol.~33, no.~2, pp. 443--445, 1985.

\bibitem{bulat2017far}
A.~Bulat and G.~Tzimiropoulos, ``How far are we from solving the 2d \& 3d face alignment problem?(and a dataset of 230,000 3d facial landmarks),'' in \emph{ICCV}, 2017, pp. 1021--1030.

\bibitem{shi2022avhubert}
B.~Shi, W.-N. Hsu, K.~Lakhotia, and A.~Mohamed, ``Learning audio-visual speech representation by masked multimodal cluster prediction,'' \emph{arXiv preprint arXiv:2201.02184}, 2022.

\bibitem{liu2019adversarial}
A.~H. Liu, H.-y. Lee, and L.-s. Lee, ``Adversarial training of end-to-end speech recognition using a criticizing language model,'' in \emph{ICASSP 2019-2019 IEEE International Conference on Acoustics, Speech and Signal Processing (ICASSP)}.\hskip 1em plus 0.5em minus 0.4em\relax IEEE, 2019, pp. 6176--6180.

\bibitem{defossez2022high}
A.~D{\'e}fossez, J.~Copet, G.~Synnaeve, and Y.~Adi, ``High fidelity neural audio compression,'' \emph{arXiv preprint arXiv:2210.13438}, 2022.

\bibitem{vasuki2006review}
A.~Vasuki and P.~Vanathi, ``A review of vector quantization techniques,'' \emph{IEEE Potentials}, vol.~25, no.~4, pp. 39--47, 2006.

\bibitem{wan2018generalized}
L.~Wan, Q.~Wang, A.~Papir, and I.~L. Moreno, ``Generalized end-to-end loss for speaker verification,'' in \emph{ICASSP 2018}.\hskip 1em plus 0.5em minus 0.4em\relax IEEE, 2018, pp. 4879--4883.

\bibitem{kim2008dysarthric}
H.~Kim, M.~Hasegawa-Johnson, A.~Perlman, J.~Gunderson, T.~S. Huang, K.~Watkin, and S.~Frame, ``Dysarthric speech database for universal access research,'' in \emph{Ninth Annual Conference of the International Speech Communication Association}, 2008.

\bibitem{veaux2016superseded}
C.~Veaux, J.~Yamagishi, K.~MacDonald \emph{et~al.}, ``Superseded-cstr vctk corpus: English multi-speaker corpus for cstr voice cloning toolkit,'' 2016.

\bibitem{zen2019libritts}
H.~Zen, V.~Dang, R.~Clark, Y.~Zhang, R.~J. Weiss, Y.~Jia, Z.~Chen, and Y.~Wu, ``Libritts: A corpus derived from librispeech for text-to-speech,'' \emph{Interspeech 2019}, 2019.

\bibitem{radford2023robust}
A.~Radford, J.~W. Kim, T.~Xu, G.~Brockman, C.~McLeavey, and I.~Sutskever, ``Robust speech recognition via large-scale weak supervision,'' in \emph{ICML}.\hskip 1em plus 0.5em minus 0.4em\relax PMLR, 2023, pp. 28\,492--28\,518.

\end{thebibliography}

\end{document}